\documentclass[baaa]{baaa}

\usepackage[pdftex]{hyperref}
\usepackage{subfigure}
\usepackage{natbib}
\usepackage{helvet,soul}
\usepackage[font=small]{caption}

\contriblanguage{1}

\contribtype{1}

\thematicarea{7}

\received{\ldots}
\accepted{\ldots}
\title{The new era of Lyman alpha emitters (LAEs): \\
Efficiency in the selection criteria}

\titlerunning{LAEs: Efficiency in selection criteria}
\author{
P. Layana-Astudillo\inst{1}, I. Laferte-Urrutia\inst{1}, L. Guaita\inst{2}, C. Artale\inst{2}, J. Magaña \inst{1}, E. Gawiser\inst{3,4}, P. Troncoso-Iribarren\inst{1}, K. Lee\inst{5} \& N. Firestone\inst{3}}

\authorrunning{Layana-Astudillo et al.}

\contact{paulette.layana@ucen.cl}

\institute{
Facultad de Ingeniería y Arquitectura, Universidad Central de Chile, Chile
\and
Instituto de Astrofísica, Universidad Andrés Bello, Chile
\and
Department of Physics and Astronomy, Rutgers, the State University of New Jersey, EE.UU.
\and
School of Natural Sciences, Institute for Advanced Study, EE.UU.
\and
Department of Physics and Astronomy, Purdue University, EE.UU.
}

\resumen{El sondeo ODIN ha detectado miles de galaxias emisoras de Lyman-alfa en siete campos del cielo, cubriendo casi cien grados cuadrados. Uno de ellos es el campo SHELA, observado con DECam en el telescopio Blanco, alcanzando profundidades de $\sim$25.3 (AB) mag en la{\bf s} bandas angostas N501/N419 y $\sim$26.3 (AB) mag en las bandas anchas \textit{g} y \textit{r}, sobre un área de quince grados cuadrados. En este trabajo analizamos la eficiencia del criterio de doble continuo en función de la profundidad de las imágenes de banda ancha. Aplicado a las observaciones N501/N419, encontramos que cuando la banda ancha es aproximadamente una magnitud más profunda que la banda angosta, se recupera cerca del 80\% de los LAEs, frente a $\sim$20\% cuando ambas poseen igual profundidad. Este resultado indica que, para un tiempo de observación fijo, la estrategia óptima consiste en obtener bandas anchas aproximadamente una magnitud más profunda que la banda angosta, ya que mayores profundidades entregan mejoras progresivamente menores en la fracción de recuperación.}

\abstract{The ODIN survey has detected thousands of Lyman-alpha emitting galaxies across seven sky fields, covering nearly one hundred square degrees. One of these regions is the SHELA field, observed with DECam on the Blanco Telescope, reaching depths of $\sim$25.3 (AB) mag in the N501/N419 narrowbands and $\sim$26.3 (AB) mag in the \textit{g} and \textit{r} broadbands over fifteen square degrees. In this work, we analyze the efficiency of the dual-continuum selection criterion as a function of broadband depth. Applied to the N501/N419 observations, we find that when the broadband data are roughly one magnitude deeper than the narrowband, nearly 80\% of the LAEs are recovered, compared to only $\sim$20\% when both reach the same depth. This result shows that, for a fixed observing budget, the optimal strategy is to obtain broadband images approximately one magnitude deeper than the narrowband, since deeper exposures yield progressively smaller gains in the recovery fraction.
}


\begin{document}

\maketitle

\section{Introduction}
\label{sec:intro}

The Lyman-$\alpha$ (Ly$\alpha$, $\lambda_{\mathrm{rest}} = 1216$\,\AA) emission line, arising from the $n=2 \rightarrow 1$ transition of neutral hydrogen, is a powerful tool for detecting faint{\bf,} high-redshift galaxies ($z>2$) through optical and near-infrared imaging \citep{Dijkstra_2014, 2020ARA&A..58..617O}. 
Ly$\alpha$ photons originate from recombination powered by massive stars or active galactic nuclei and can scatter through the circumgalactic and intergalactic medium, making them key probes of galaxy evolution and cosmic reionization.

Galaxies with strong Ly$\alpha$ emission, known as Lyman-$\alpha$ emitters (LAEs), are typically young, low-mass, star-forming systems with low dust content and rest-frame equivalent widths ($EW_0$) commonly exceeding 20~\AA.
\citep{Gawiser_2006, Schaerer_2009, Nakajima_2011, Ouchi_2013}.
% This emission is detected in narrowbands and the adjacent broadbands records the contrast, .}

Several wide-field surveys have targeted LAEs by exploiting their prominent narrowband excess relative to broadband filters. 
These include the Systematic Identification of LAEs for Visible Exploration and Reionization Research Using Subaru HSC (SILVERRUSH) \citep{2018PASJ...70S..13O}, 
Lyman Alpha Galaxies in the Epoch of Reionization (LAGER) \citep{Zheng_2016}, and One-hundred-deg$^2$ DECam Imaging in Narrowbands (ODIN) \citep{lee2023onehundreddeg2decamimagingnarrowbands}. 
They substantially increased the number of known LAEs by sampling larger areas of the sky, deeper imaging in narrow and broadband filters, and multi-narrowband strategies probing multiple redshifts \citep{2020ARA&A..58..617O} compared to the first efforts (Large Area Lyman Alpha Survey (LALA) \citep{rhoads2001lala} and the Multiwavelength Survey by Yale–Chile (MUSYC) \citep{2008ApJ...681.1099B}).
Yet, variations in the width and depth of the narrowband, the broadband, and the difference between both might introduce selection biases in the detected LAE population and their inferred properties, motivating a quantitative assessment of the depth balance between narrow and broadband imaging \citep{deDiego13}.

The identification of protoclusters and large-scale structures (LSS) has been boosted by the growth of LAE samples. For instance, ODIN has revealed protoclusters and extended Ly$\alpha$ blobs (LABs) tracing overdense regions along cosmic filaments \citep{2024ApJ...977..119R,2023ApJ...951..119R}, consistent with predictions from simulations such as \textit{IllustrisTNG} and \textit{Horizon Run 5} \citep{andrews,im2024testing}.

In this work, we quantify the detection efficiency of LAEs as a function of broadband depth using ODIN data in the SHELA field \citep{papovich16}. 
Our analysis provides practical guidelines for optimizing exposure-time allocation and improving the completeness of wide-field photometric LAE searches 
at $z=2.4, 3.1$. Its dependence with the narrowband depth and width is out of the scope of this article. 

\vspace{-4.6mm}
\section{Data}
\label{sec:data}

\subsection{ODIN: One-hundred-deg$^2$ DECam Imaging in Narrowbands}

\begin{figure*}%[!t]
\centering
\includegraphics[width=0.8\textwidth]{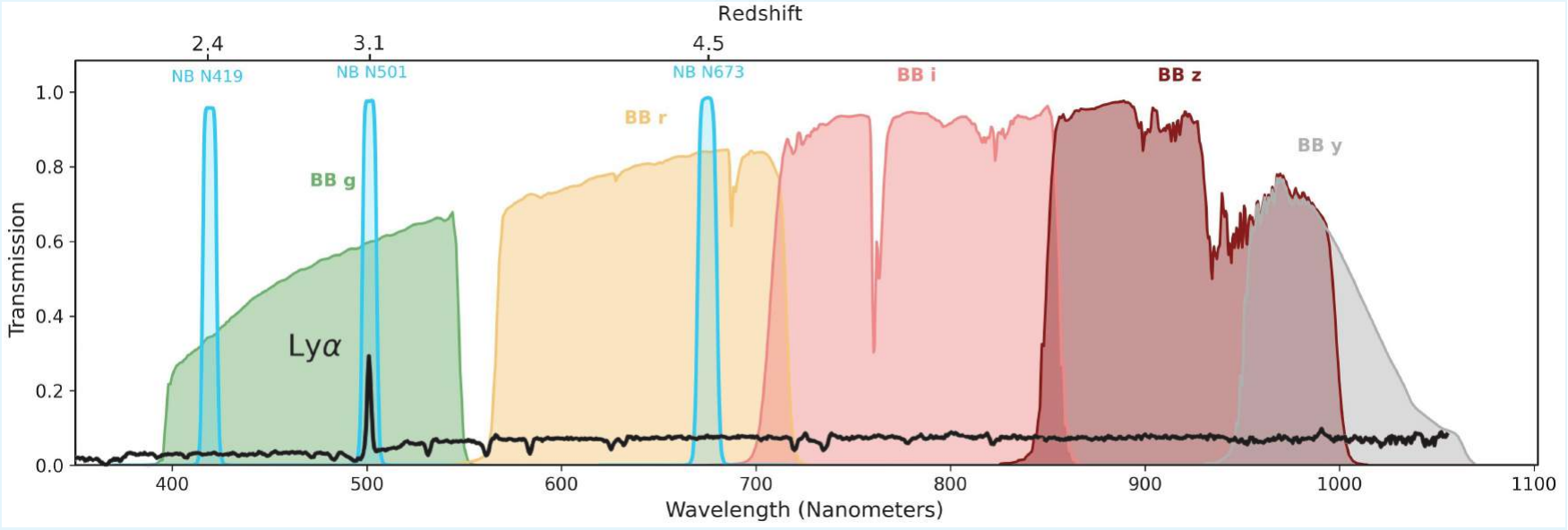}
\caption{
Transmission curves of the ODIN narrowband filters designed for DECam (blue shaded regions; from left to right: N419, N501, and N673), together with the corresponding broadband filters $g$, $r$, $i$, $z$, and $y$ obtained from the DECam DR1 Standard Bandpasses \citep{2018ApJS..239...18A}. 
The horizontal axis indicates wavelength in nanometers. 
These imaging data are combined to isolate redshifted Ly$\alpha$ emission falling into the narrowband filters, as illustrated by the template LAE spectrum (black) at $z = 3.1$. 
The corresponding Ly$\alpha$ redshift range for each narrowband is shown along the top axis. 
\textit{Adapted from} \citet{lee2023onehundreddeg2decamimagingnarrowbands}.}
\label{fig:bbynb}
\end{figure*}

The \textit{ODIN} survey studies the role of LSS environments in galaxy formation during the \textit{Cosmic Noon}. Observations began in early 2020 to map LSS through LAEs as tracers \citep{lee2023onehundreddeg2decamimagingnarrowbands}. ODIN uses the Dark Energy Camera (DECam) on the Blanco Telescope at CTIO with three custom narrowband (NB) filters: N419, N501, and N673, centered at $419~\mathrm{nm}$, $501~\mathrm{nm}$, and $673~\mathrm{nm}$, with full widths at half maximum (FWHM) of $7.5~\mathrm{nm}$, $7.6~\mathrm{nm}$, and $10.0~\mathrm{nm}$, respectively. Corresponding to redshifts $z \simeq 2.4,\; 3.1,$ and $4.5$. The transmission curves of the NB and DECam broadband ($grizy$) filters are shown in Fig. \ref{fig:bbynb}. 

LAE selection is based on a rest-frame equivalent width threshold EW$_0 > 20~\mathrm{\AA}$.This threshold minimizes the contamination of low-redshift emission lines such as [O\,II] or [O\,III], commonly referred to as \textit{interlopers}. 
In its first six years, ODIN has identified more than $16\,000$ LAE only in the COSMOS field \citep{2025ApJ...995..126N}. Adding the LAEs detected in the other six fields will enable statistically robust analyzes of the cosmic web.

\vspace{-3mm}
\subsection{The Spitzer-HETDEX Exploratory Large Area Survey (SHELA) field }

The \textit{SHELA} field \citep{papovich16} is one of ODIN’s seven survey regions.
Although not part of any ultra-deep imaging program, ODIN has progressively improved the broadband depth across SHELA to achieve uniform sensitivity, prioritizing the $g$ and $r$-bands and ensuring consistent LAE selection.
This field includes HETDEX integral-field spectroscopy and Dark Energy Spectroscopic Instrument (DESI) used to confirm the brightest detections.
The field covers eight DECam pointings, namely P1 to P8.
The latest observations, corresponding to pointings P3 and P4, were completed during the 2025B semester, reaching a comparable depth to the previously acquired data. The LAE selection for these pointings is still in progress and will complete the ODIN–SHELA coverage.

\vspace{-2mm}
\section{Methodology}
\label{sec:methodology}

We analyze six ODIN–SHELA pointings, namely the subfields P12 (P1 and P2), P56 (P5 and P6), and P78 (P7 and P8), which together cover an effective area of $\sim$$15~\mathrm{deg}^2$. 
This subset represents the most homogeneous region currently available within SHELA for LAE detection at $z=2.4, 3.1$, probed through the N501/N419 narrowband filters.
%This analysis focuses on LAEs detected with the N501 narrowband filter ($z \simeq 3.1$) in six ODIN--SHELA subfields.}

\vspace{-4mm}
\subsection{Photometry}

Archival DECam and CNTAC data were reprocessed using the \texttt{Community Pipeline} (CP), reaching typical $5\sigma$ depths of $g/r\simeq26.3$ and N419/N501$\simeq25.3$\,AB mag for P12, P56, and P78.
The data were processed following the procedure described in Section 2 of \citet{Firestone_2024}. Photometry was performed using \texttt{SExtractor} in dual-image mode, using the narrowband for detection and matched-aperture fluxes in the broadband images. A fixed  $2~\mathrm{arcsec}$ aperture was used, following the ODIN pipeline. Fluxes were corrected for Galactic extinction using the reddening maps of \citet{Schlegel1998}, recalibrated by \citet{Schlafly2011}, and the extinction law of \citet{Fitzpatrick1999}.

\vspace{-4mm}
\subsection{Selecting LAEs}

LAE candidates were selected using the hybrid-weighted double-broadband continuum criteria described by \citet{Firestone_2024}, based on narrowband excess relative to a weighted combination of the $g$ and $r$ bands that trace the continuum near Ly$\alpha$. The selection is based on the color-excess criterion:

\begin{equation}
(BB - NB) > \Delta_{\mathrm{min}},
\end{equation}

where $BB = -0.438\,g + 1.438\,r, (0.856\,g + 0.144\,r) $ represents the weighted broadband continuum estimate and $\Delta_{\mathrm{min}} = 0.71(0.83)~\mathrm{mag}$ for the N419(N501) filter, respectively.
ODIN candidates confirmed spectroscopically by DESI indicate a negligible contamination of [O\,II] and [O\,III] emitters \citep{2026arXiv260309905P}, reporting confirmation rates of (93, 96, 92)\% at $z=(2.4, 3.1, 4.5)$. 
The minimum EWs detection is determined by the NB depth. In this work we only vary the BB depth, so we do not study a variation of the number of interlopers as a function of BB depth or narrowband width. Since only the BB depth is varied here, no strong dependence on $\Delta(\mathrm{BB}-\mathrm{NB})$ is expected, and contamination should not significantly affect the recovery trends.

\vspace{-4mm}
\subsection{Recovery Fraction}

%The LAE selection depends on the relative depth of broad and narrowband imaging
LAE selection depends on BB–NB depth, variations in broadband depth can cause genuine emitters to be missed. The recovery fraction estimates sample completeness under different depth configurations. The recovery fraction is defined as

\begin{equation}
f_{\mathrm{rec}} =
\frac{N_{\mathrm{rec}}}{N_{\mathrm{depth}}},
\end{equation}

where $N_{\mathrm{rec}}$ is the number of LAE candidates recovered at a given depth and $N_{\mathrm{depth}}$ corresponds to the total number detected in the deepest configuration, in this work we use $\Delta(\mathrm{BB}-\mathrm{NB}) = 2.5$. To evaluate its dependence on broadband depth, we combined ODIN simulations with real ODIN--SHELA observations.

\vspace{-4mm}
\subsubsection{Simulations of Observational Data}

The simulation framework is based on deep imaging in the COSMOS field, where the $g$-band data are $\sim 2.5$ magnitudes deeper than the narrowband. The broadband depth was artificially degraded by inflating the flux uncertainties with additional Gaussian noise while preserving the original point-spread function. LAE candidates were then re-selected under each depth configuration, ensuring that simulated galaxies follow the luminosity distribution of ODIN LAEs. This approach allows us to quantify how the recovery fraction varies as a function of the relative depth $\Delta = \mathrm{BB} - \mathrm{NB}$.

\vspace{-4mm}
\subsubsection{Observational Data}
To enable a direct comparison with the simulations performed in the previous section, we constructed magnitude-limited subsamples by progressively imposing brighter magnitude cuts on the broadband catalogs. Each configuration mimics a shallower depth while preserving source positions and photometry. The catalogs are complete down to the limiting magnitude of the deepest configuration, ensuring that the reference sample ($N_{\mathrm{depth}}$) represents the most complete LAE set.

\vspace{-3mm}
\subsection{LSS Distribution}
To study the spatial distribution of LAEs, Gaussian kernel smoothing was applied to their projected sky positions, generating surface-density maps for the six SHELA sub-fields. Overdensities were defined as regions exceeding $2\sigma$ and $3\sigma$ above the mean field density.

%Surface-density maps were constructed using Gaussian kernel smoothing, and overdensities were defined as regions exceeding $2\sigma$, $3\sigma$, and $4\sigma$ above the mean field density.}

\vspace{-5mm}
\section{Results}
\label{sec:results}
\begin{figure}[h!]
\centering
\includegraphics[width=0.45\textwidth]{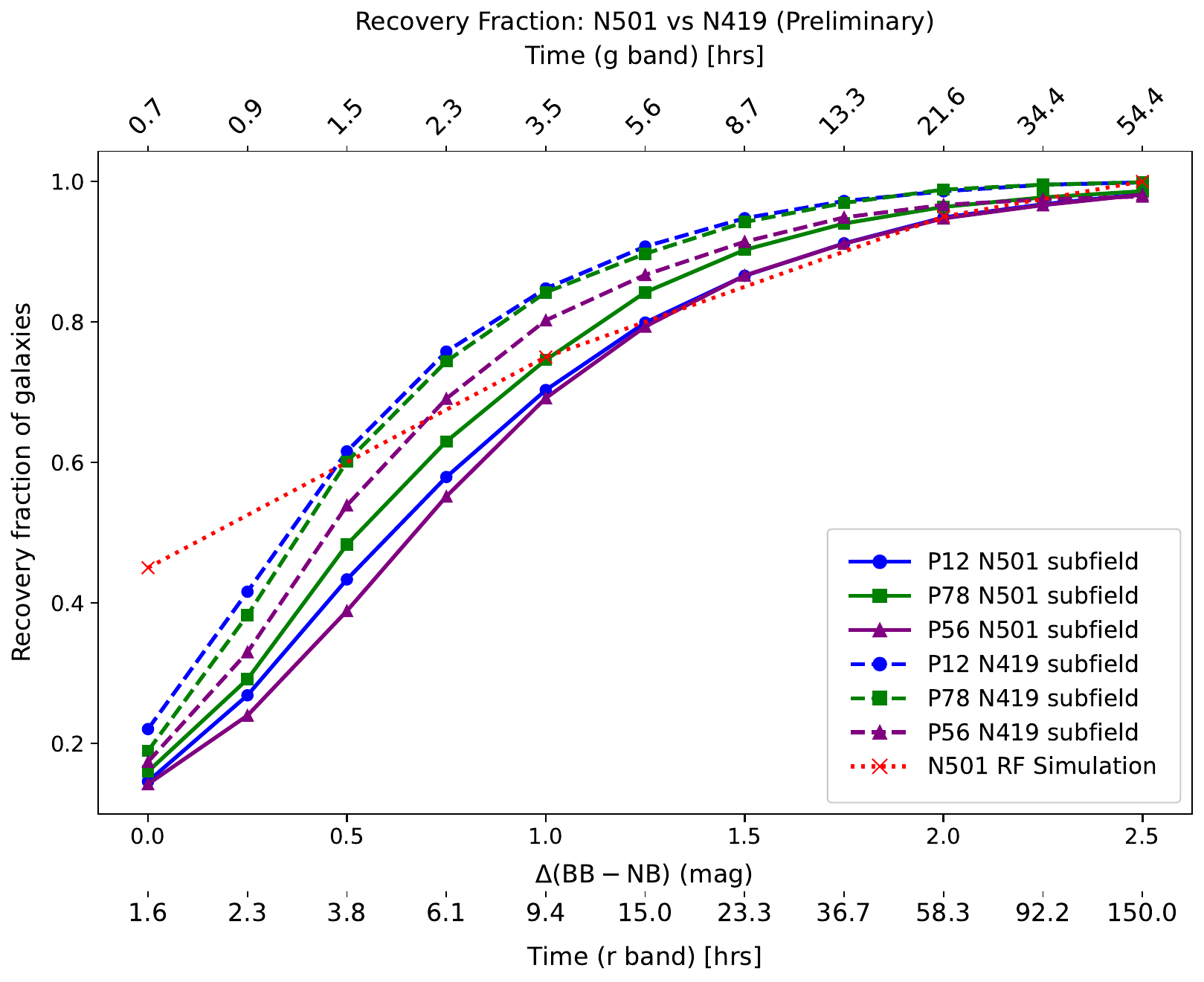}
\caption{Recovery fraction of LAE candidates as a function of the broadband–narrowband depth difference, $\Delta(\mathrm{BB}-\mathrm{NB})$, for the ODIN–SHELA pointings P12, P56, and P78. 
Solid lines correspond to the N501 selection, while dashed lines represent the N419 selection, with consistent color coding for each subfield: P12 (blue circles), P56 (purple triangles), and P78 (green squares). 
The red dotted line shows the predicted recovery fraction from the ODIN collaboration simulation for N501. 
Top and bottom axes indicate the corresponding exposure times required in the $g$ and $r$ bands under dark and grey time conditions, respectively. 
The N419 results are shown for comparison and should be considered preliminary.}
\label{fig:recovery_fraction}
\end{figure}

Figure~\ref{fig:recovery_fraction} shows the recovery fraction of LAEs as a function of $\Delta(\mathrm{BB}-\mathrm{NB})$.
The observational results of the ODIN--SHELA subfields P12, P56, and P78 are shown in blue, green, and purple dashed and filled lines for N419, and N501, respectively. 
The red dotted line shows the N501 simulation,
the observational results follow a consistent trend with it. %as no simulation is currently available for N419;

\vspace{-3mm}
The recovery fraction increases steadily with broadband depth for both simulation and observational results. 
At $\Delta \approx 0$, only $\sim 40\%$ ($\sim 20\%$) of LAEs are recovered in the simulation (observations), highlighting the impact of broadband depth.
For observations, $f_{rec}$ increases from $\sim40\%$ at $\Delta \approx 0$ up to $\sim70$--$80\%$ at $\Delta \sim 1$, with smaller gains at larger depths.
The upper axis indicates the $g$-band exposure time (dark time) required to reach a given depth, while the lower axis shows the corresponding $r$-band time (gray time). Both are calculated using the DECam Exposure Time Calculator (ETC).

\begin{figure*}[h!]
\centering
\includegraphics[width=0.92\linewidth]{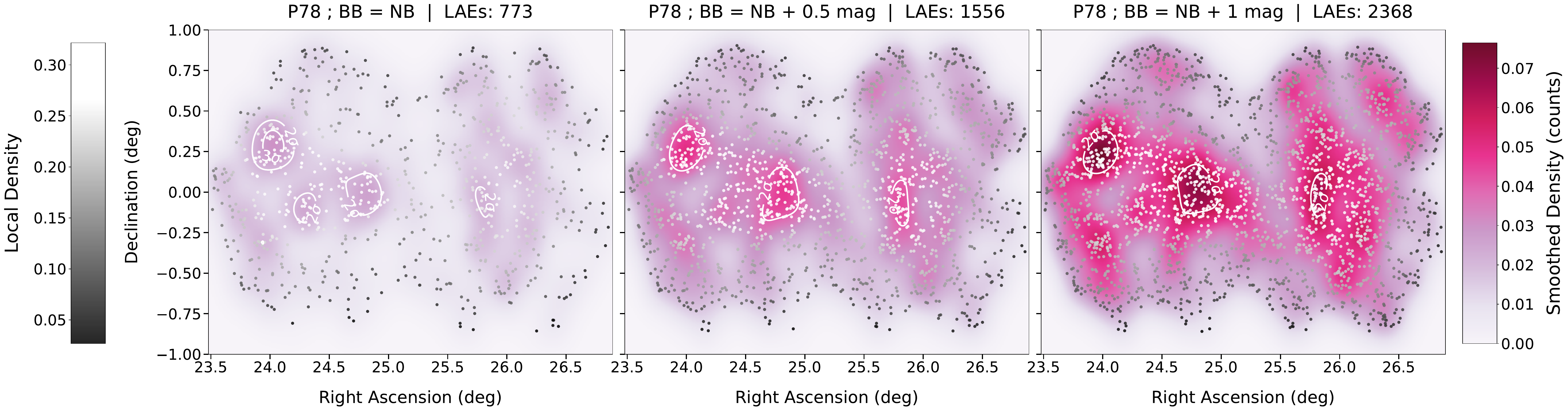}
\caption{
Smoothed spatial distribution of LAE candidates in the ODIN--SHELA subfield P78 for three broadband depths relative to the narrowband.
\emph{Left panel:} $\mathrm{BB} = \mathrm{NB}$. 
\emph{Central panel:} $\mathrm{BB} = \mathrm{NB} + 0.5$ mag. 
\emph{Right panel:} $\mathrm{BB} = \mathrm{NB} + 1$ mag. 
The Gaussian-smoothed background represents the surface density of LAEs, while contours mark overdensities at $2\sigma$ and $3\sigma$ above the mean field density. 
Colored points indicate individual LAEs, with their color encoding the local density. 
The number of detected candidates in each configuration is reported above each panel.
}
\label{fig:p78_depth}
\end{figure*}

Figure~\ref{fig:p78_depth} illustrates how this effect impacts the spatial distribution of LAEs in the P78 subfield for the N501 filter. For $\mathrm{BB}=\mathrm{NB}$, the large-scale structure is barely visible; at $\mathrm{BB}=\mathrm{NB}+0.5$ mag, overdense regions begin to emerge; and at $\mathrm{BB}=\mathrm{NB}+1$ mag, the main structures are clearly traced, consistent with the $\sim75$--$80\%$ recovery fraction.
%%%%%%%%%%%%%%%%%%%%%%%%%%%%%%%%%%%%%%%%%%%%%%%%%%%%%%%%%%%%%%%%%%%%%%
% Para figuras de dos columnas use \begin{figure*} ... \end{figure*}         %
%%%%%%%%%%%%%%%%%%%%%%%%%%%%%%%%%%%%%%%%%%%%%%%%%%%%%%%%%%%%%%%%%%%%%%%%%%%%%%

\vspace{-3mm}
\section{Discussion and Conclusions}
\label{sec:discussion}

We analyzed the efficiency of LAE detection in the ODIN--SHELA field as a function of broadband depth, following the selection method of \citet{Firestone_2024}, and quantified how the recovery fraction depends on the relative broadband–narrowband depth.

Our results show that LAE completeness is strongly driven by this depth balance. When the broadband images are $\sim1$ mag deeper than the narrowband, the recovery fraction increases from $\sim20\%$ to nearly $80\%$, in agreement with the ODIN simulations, demonstrating that relative depth is a key factor governing photometric emission-line selection.

This result is derived for N419/N501 ($z\sim2.4,3.1$). The preliminary N419 behavior is consistent with the same overall trend, although confirmation requires a full multi-filter analysis. The exact optimal balance may vary for other filters, such as N673, and for different LAE luminosities and surface-brightness distributions.

This behavior can be understood as the selection being primarily set by the narrowband excess threshold, while the broadband depth mainly determines whether genuine LAEs are recovered. As a result, completeness increases up to $\Delta \sim 1$, while contamination is not expected to vary strongly with BB depth alone.

From an observational perspective, these results suggest that an efficient strategy is to obtain broadband data roughly one magnitude deeper than the narrowband images. Using the DECam ETC, we estimate that this configuration requires comparable observing time in NB and BB.

For depth differences larger than $\Delta \sim 1$, the increase in recovery fraction becomes progressively smaller, indicating that most detectable LAEs are already recovered and additional exposure time yields only marginal improvements in completeness.

The spatial analysis of the P78 field further illustrates the impact of broadband depth on large-scale structure reconstruction: when $\mathrm{BB}=\mathrm{NB}$, only a small fraction of LAEs is recovered, limiting the visibility of the structure; whereas $\mathrm{BB}=\mathrm{NB}+1$ mag reveals most filamentary features traced by LAEs.

%This analysis focuses on the N501 filter at $z\sim3.1$; quantitative values may differ for other filters (e.g., N419 at $z\sim2.4$ or N673 at $z\sim4.5$), where intrinsic LAE properties and observing conditions vary. Although the optimal BB–NB balance may depend on luminosity and surface-brightness distributions, the general trend highlighting the importance of relative depth is expected to remain valid for similar survey strategies.

%\vspace{-3mm}
\begin{acknowledgement}
PL acknowledges support from the Proyecto Puente CIPPTE202501-UCEN 2025. 
PTI acknowledges the use of the GÜINA funded by EQM200216 and the support provided by the Agencia Nacional de Investigación y Desarrollo (ANID) through its program Fomento a la Vinculación Internacional (FOVI) number 240098.
JM acknowledges the support of the project Fondecyt Regular No. 1240514.
\end{acknowledgement}

%%%%%%%%%%%%%%%%%%%%%%%%%%%%%%%%%%%%%%%%%%%%%%%%%%%%%%%%%%%%%%%%%%%%%%%%%%%%%%
%  ******************* Bibliografía / Bibliography ************************  %
%                                                                            %
%  -Ver en la sección 3 "Bibliografía" para mas información.                 %
%  -Debe usarse BIBTEX.                                                      %
%  -NO MODIFIQUE las líneas de la bibliografía, salvo el nombre del archivo  %
%   BIBTEX con la lista de citas (sin la extensión .BIB).                    %
%                                                                            %
%  -BIBTEX must be used.                                                     %
%  -Please DO NOT modify the following lines, except the name of the BIBTEX  %
%  file (without the .BIB extension).                                       %
%%%%%%%%%%%%%%%%%%%%%%%%%%%%%%%%%%%%%%%%%%%%%%%%%%%%%%%%%%%%%%%%%%%%%%%%%%%%%% 

\bibliographystyle{baaa}
\small
\bibliography{bibliografia}
\end{document}